\documentclass[12pt,a4paper]{article}
\usepackage[margin=1in]{geometry}
\usepackage{amsmath,amssymb,mathrsfs}
\usepackage{graphicx}
\usepackage{booktabs}
\usepackage{algorithm,algpseudocode}
\usepackage{xcolor}
\usepackage{multirow}
\usepackage{siunitx}
\usepackage{caption,subcaption}
\usepackage{authblk}  
\usepackage{hyperref}
\hypersetup{colorlinks=true, linkcolor=blue, citecolor=blue, urlcolor=blue}
\usepackage[numbers]{natbib} 
\newcommand{\vect}[1]{\mathbf{#1}}
\newcommand{\mat}[1]{\mathbf{#1}}
\newcommand{\expected}[1]{\mathbb{E}\left[#1\right]}

\begin{document}

\title{\Large \bf Micro-Doppler Energy-Based Robust Multi-Target Vital Signs Monitoring Using 77-GHz FMCW Radar with Spatiotemporal Adaptive Processing}
\author[1]{Chenxing Tan \thanks{\href{mailto:3424611356@qq.com}{3424611356@qq.com}}}
\author[1]{Yuguan Hou$^*$ \thanks{\href{mailto:yuguanhou@hit.edu.cn}{yuguanhou@hit.edu.cn}}}  
\author[1]{Hao Wang \thanks{\href{2922068495@qq.com}{2922068495@qq.com}}}  
\author[1]{Zhonghao Yuan \thanks{\href{mailto:yuanatp@qq.com}{yuanatp@qq.com}}}
\affil[1]{School of Electronics and Information Engineering, Harbin Institute of Technology, Harbin 150001, China\\
$^*$Corresponding author}  
\date{}
\maketitle

\begin{abstract}
This paper presents a novel micro-Doppler energy-based framework for robust multi-target vital signs monitoring using 77-GHz Frequency-Modulated Continuous-Wave (FMCW) radar. Unlike conventional phase-based methods that are susceptible to environmental noise, random body movements, and stringent calibration requirements, our approach exploits the energy variations in radar returns induced by cardiopulmonary activities. The proposed system integrates a comprehensive processing pipeline including space-time adaptive processing (STAP) for target detection and tracking, MUSIC algorithm for high-resolution angle estimation, and an innovative adaptive spectral filtering technique for vital signs extraction. We establish a rigorous mathematical framework that formalizes the relationship between micro-Doppler energy variations and physiological activities, enabling robust separation of closely spaced targets. The key innovation lies in the micro-Doppler energy extraction methodology that provides inherent robustness to phase noise and motion artifacts. Experimental results using millimeter-wave radar datasets demonstrate that the system can accurately detect and separate vital signs of up to four targets within \SI{5}{\meter} range, achieving mean absolute errors of \SI{1.2}beats per minute and \SI{2.3} beats per minute for respiration and heart rates, respectively. The proposed approach demonstrates superior performance compared to traditional phase-based methods, particularly in challenging multi-target scenarios with environmental noise and subject movement.
\end{abstract}

\vspace{0.5cm} 

\section{Introduction}
Non-contact vital signs monitoring has emerged as a transformative technology for healthcare applications, security screening, and human-computer interaction \cite{zhang2021noncontact, lewandowska2011measuring, bush2016measuring}. Traditional radar-based approaches predominantly rely on phase variations induced by chest wall movements to extract physiological signals \cite{kayani2020pulse, atumann2018radiating}. While phase-based methods can achieve high accuracy under controlled conditions, they face fundamental limitations in practical scenarios: sensitivity to environmental noise, vulnerability to random body movements, requirement for precise calibration, and difficulty in handling multiple targets simultaneously \cite{vinci201224, diraco2017radar}.

Micro-Doppler energy-based approaches present a paradigm shift by analyzing the energy variations in radar returns caused by cardiopulmonary activities. These methods leverage the physical principle that physiological movements modulate the radar cross-section (RCS) and distribution of reflected energy, providing a more robust signature compared to phase variations \cite{chian2022vital, naiabadham2016estimation}. Frequency-Modulated Continuous-Wave (FMCW) radar operating in millimeter-wave bands (particularly \SI{77}{\giga\hertz}) provides an ideal platform for implementing micro-Doppler energy-based vital signs monitoring due to its high range resolution, penetration capability, and ability to track multiple targets \cite{mercuri2017frequency, mercuri2018direct}.

Despite these advantages, simultaneous monitoring of multiple subjects remains challenging due to several fundamental factors: (1) interference and coupling between closely spaced targets; (2) limited spatial resolution of conventional radar systems; (3) difficulty in separating physiological signals from motion artifacts and environmental noise; (4) computational complexity of real-time processing for multiple targets; and (5) the need for robust association between detected targets and their physiological signals \cite{laurealager2019coherent, fan2023radiofrequency}.

In this paper, we propose a comprehensive micro-Doppler energy-based framework for robust multi-target vital signs monitoring using \SI{77}{\giga\hertz} FMCW radar. Our main contributions include:

\begin{enumerate}
\item A rigorous mathematical formulation of micro-Doppler energy extraction that establishes the fundamental relationship between radar energy variations and physiological activities;
\item An integrated space-time adaptive processing (STAP) approach combined with MUSIC algorithm for high-resolution target detection and tracking;
\item A multi-stage signal processing pipeline incorporating wavelet denoising, adaptive spectral filtering, and harmonic suppression for robust vital signs extraction;
\item A theoretical analysis comparing micro-Doppler energy and phase-based methods, demonstrating the fundamental advantages of the energy-based approach;
\item Extensive experimental validation demonstrating superior performance over traditional phase-based methods in multi-target scenarios.
\end{enumerate}

The remainder of this paper is organized as follows: Section 2 reviews related work on radar-based vital signs monitoring. Section 3 describes the system overview and radar signal model with emphasis on micro-Doppler energy formulation. Section 4 details our proposed methodology with comprehensive mathematical derivations. Section 5 presents experimental results and discussion. Finally, Section 6 concludes the paper and suggests future research directions.

\section{Related Work}
Radar-based vital signs monitoring has evolved substantially over the past decade, progressing from basic Doppler radar systems to sophisticated FMCW and MIMO radar architectures. Early research focused on single-target detection using continuous-wave Doppler radar systems \cite{kayani2020pulse, atumann2018radiating}. These systems could accurately detect respiration and heart rates but were fundamentally limited to monitoring one subject at a time and relied exclusively on phase information, making them vulnerable to environmental disturbances.

The development of FMCW radar enabled range resolution, providing limited multi-target capability by separating targets based on distance \cite{vinci201224, diraco2017radar}. Multiple-input multiple-output (MIMO) radar systems further enhanced angular resolution through virtual array techniques, enabling better spatial separation of targets \cite{chian2022vital, naiabadham2016estimation}. However, these systems often require complex hardware configurations and sophisticated signal processing algorithms. Conventional beamforming techniques have been employed to steer radar beams toward specific targets, but they suffer from limited resolution when targets are closely spaced and require accurate angle-of-arrival estimation \cite{naiabadham2016estimation, mercuri2017frequency}.

Recent advances incorporate machine learning techniques for vital signs monitoring. Deep learning approaches can learn complex patterns from radar data but require large annotated datasets for training, which are often difficult to obtain for physiological signals \cite{gu2010instrument, kazemi2016vital}. Unsupervised learning methods like Independent Component Analysis (ICA) have been used to separate mixed signals but can be computationally intensive and sensitive to initialization parameters \cite{zhao2017noncontact, tang2019corcoupled}.

Phase-based methods, while achieving high accuracy in controlled environments, face fundamental limitations in practical applications. The phase of radar signals is highly sensitive to environmental changes, requiring frequent calibration. Small body movements can cause phase wrapping issues, and the method struggles with multiple targets due to phase mixing \cite{chen2008implementation, yang2014doppler}. The fundamental limitation stems from the relationship:

\begin{equation}
\phi(t) = \frac{4\pi R(t)}{\lambda} \mod 2\pi
\end{equation}

where $\phi(t)$ is the measured phase, $R(t)$ is the time-varying distance to the target, and $\lambda$ is the wavelength. This modulo $2\pi$ operation causes ambiguity when $R(t) > \lambda/2$, leading to phase wrapping that disrupts vital signs estimation.

Micro-Doppler-based approaches have gained attention for their robustness to environmental factors. These methods analyze the frequency shifts and energy variations caused by micro-motions, providing more stable features for physiological monitoring \cite{wang2014noncontact, oh2021development}. However, existing micro-Doppler methods have primarily focused on single-target scenarios or require complex hardware setups.

Compared to existing approaches, our method offers several distinct advantages: (1) it uses micro-Doppler energy instead of phase information, providing better robustness to noise and environmental interference; (2) it incorporates a multi-stage processing pipeline that handles various signal quality issues; (3) it does not require extensive training data; (4) it provides accurate vital signs estimation for multiple targets simultaneously; and (5) it operates in real-time on standard hardware.

\section{System Overview and Radar Signal Model}
\subsection{FMCW Radar Fundamentals}
FMCW radar transmits chirp signals with frequency linearly increasing over time. The transmitted signal can be expressed as:
\begin{equation}
s_T(t) = A_T \exp\left[j2\pi\left(f_c t + \frac{1}{2}\alpha t^2\right)\right], \quad 0 \leq t \leq T_c
\end{equation}
where $A_T$ is the amplitude, $f_c$ is the carrier frequency, $\alpha = B/T_c$ is the chirp rate, $B$ is the bandwidth, and $T_c$ is the chirp duration.

The reflected signal from a target at distance $R(t)$ is received after a time delay $\tau(t) = 2R(t)/c$:
\begin{equation}
s_R(t) = A_R \exp\left[j2\pi\left(f_c (t-\tau(t)) + \frac{1}{2}\alpha (t-\tau(t))^2\right)\right]
\end{equation}
where $A_R$ is the received signal amplitude, which depends on the radar cross-section (RCS) of the target and propagation losses.

The beat signal is obtained by mixing the transmitted and received signals:
\begin{align}
s_b(t) &= s_T(t) \cdot s_R^*(t) \\
&= A_b \exp\left[j2\pi\left(f_c\tau(t) + \alpha\tau(t)t - \frac{1}{2}\alpha\tau^2(t)\right)\right]
\end{align}
Typically, the last term ($\frac{1}{2}\alpha\tau^2(t)$) is negligible, and the beat signal simplifies to:
\begin{equation}
s_b(t) \approx A_b \exp\left[j2\pi\left(\frac{2\alpha R(t)}{c}t + \frac{2f_c R(t)}{c}\right)\right]
\end{equation}

\subsection{Micro-Doppler Energy Model}
Unlike conventional phase-based methods that rely on the phase term $\frac{4\pi R(t)}{\lambda}$, our approach analyzes the energy variations caused by micro-motions. The received signal energy depends on the radar cross-section (RCS) of the target, which is modulated by physiological activities.

For a point target with complex reflectivity $\Gamma(t)$, the received signal energy can be expressed as:
\begin{equation}
E_r(t) = |\Gamma(t)|^2 \cdot \frac{G_t G_r \lambda^2 \sigma}{(4\pi)^3 R^4(t)} P_t
\end{equation}
where $G_t$ and $G_r$ are the transmitter and receiver antenna gains, $\lambda$ is the wavelength, $\sigma$ is the radar cross-section, $R(t)$ is the target distance, and $P_t$ is the transmitted power.

For physiological monitoring, the time-varying component of the RCS due to cardiopulmonary activities can be modeled as:
\begin{equation}
|\Gamma(t)|^2 = \Gamma_0 + \Gamma_r \cos(2\pi f_r t + \phi_r) + \Gamma_h \cos(2\pi f_h t + \phi_h) + n_{\Gamma}(t)
\end{equation}
where $\Gamma_0$ is the static RCS component, $\Gamma_r$ and $\Gamma_h$ are the RCS modulation amplitudes for respiration and heartbeat, respectively, $f_r$ and $f_h$ are the corresponding frequencies, and $n_{\Gamma}(t)$ represents noise and interference in the RCS.

The micro-Doppler energy for a target at range bin $r_0$ and Doppler bin $d_0$ is defined as:
\begin{equation}
E(i) = \sum_{rx=1}^{N_{rx}} \sum_{r=r_0-\Delta r}^{r_0+\Delta r} \sum_{d=d_0-\Delta d}^{d_0+\Delta d} |S_{\text{doppler}}(rx, d, r, i)|^2
\end{equation}
where $S_{\text{doppler}}$ is the range-Doppler map, $N_{rx}$ is the number of receive antennas, and $\Delta r$, $\Delta d$ define the extraction region size around the target. This formulation captures the energy variations caused by physiological activities while being less sensitive to phase noise.

The micro-Doppler energy can be related to physiological activities through:
\begin{equation}
E(t) = E_0 + E_r \cos(2\pi f_r t + \phi_r) + E_h \cos(2\pi f_h t + \phi_h) + n(t)
\end{equation}
where $E_0$ is the DC component representing static reflections, $E_r$ and $E_h$ are the energy amplitudes for respiration and heartbeat, respectively, $f_r$ and $f_h$ are the corresponding frequencies, and $n(t)$ represents noise and interference.

\subsection{Theoretical Comparison: Micro-Doppler Energy vs. Phase-Based Methods}
The micro-Doppler energy approach offers several fundamental advantages over traditional phase-based methods, which can be understood through mathematical analysis.

\subsubsection{Robustness to Phase Noise}
Phase-based methods are highly sensitive to phase noise caused by environmental fluctuations and hardware imperfections. The phase measurement can be modeled as:
\begin{equation}
\phi_{\text{measured}}(t) = \frac{4\pi R(t)}{\lambda} + \phi_{\text{noise}}(t) \mod 2\pi
\end{equation}
where $\phi_{\text{noise}}(t)$ represents phase noise with variance $\sigma_{\phi}^2$. For micro-Doppler energy methods, the measurement is:
\begin{equation}
E_{\text{measured}}(t) = E(t) + n_E(t)
\end{equation}
where $n_E(t)$ is energy measurement noise with variance $\sigma_E^2$. In practical environments, $\sigma_{\phi}^2 \gg \sigma_E^2$ due to the inherent stability of energy measurements compared to phase measurements.

\subsubsection{Immunity to Phase Wrapping}
Phase wrapping occurs when the chest displacement exceeds $\lambda/4$, causing ambiguity in phase measurement:
\begin{equation}
\phi_{\text{wrap}}(t) = \frac{4\pi R(t)}{\lambda} \mod 2\pi = \frac{4\pi}{\lambda} \left(R(t) - \frac{\lambda}{2} \left\lfloor \frac{2R(t)}{\lambda} + \frac{1}{2} \right\rfloor \right)
\end{equation}
This wrapping disrupts vital signs extraction, particularly for heart rate monitoring where displacements can approach $\lambda/4$ at \SI{77}{\giga\hertz} ($\lambda \approx \SI{3.9}{\milli\meter}$). Micro-Doppler energy methods are immune to this issue as they do not rely on phase unwrapping.

\subsubsection{Superior Motion Artifact Rejection}
Random body movements affect both phase and energy measurements, but the impact differs fundamentally. For phase-based methods, body movements introduce large phase shifts:
\begin{equation}
\phi_{\text{motion}}(t) = \frac{4\pi}{\lambda} \Delta R_{\text{motion}}(t)
\end{equation}
which can overwhelm the physiological signal. For micro-Doppler energy methods, motion artifacts appear as additive noise:
\begin{equation}
E_{\text{motion}}(t) = E_{\text{physio}}(t) + E_{\text{motion}} + n(t)
\end{equation}
which can be more effectively filtered using appropriate signal processing techniques.

\subsubsection{Mathematical Formulation of Advantage}
The fundamental advantage can be quantified through the signal-to-noise ratio (SNR) of the extracted physiological signal. For phase-based methods, the SNR is:
\begin{equation}
\text{SNR}_{\text{phase}} = \frac{\left(\frac{4\pi}{\lambda} \Delta R_{\text{physio}}\right)^2}{\sigma_{\phi}^2 + \left(\frac{4\pi}{\lambda} \Delta R_{\text{motion}}\right)^2}
\end{equation}
where $\Delta R_{\text{physio}}$ is the physiological displacement, and $\Delta R_{\text{motion}}$ is the motion artifact displacement.

For micro-Doppler energy methods, the SNR is:
\begin{equation}
\text{SNR}_{\text{energy}} = \frac{\left(\Delta E_{\text{physio}}\right)^2}{\sigma_E^2 + \left(\Delta E_{\text{motion}}\right)^2}
\end{equation}

In typical scenarios, $\text{SNR}_{\text{energy}} > \text{SNR}_{\text{phase}}$ due to the smaller denominator term, confirming the theoretical advantage of the micro-Doppler energy approach.

\subsection{System Architecture}
\begin{figure}[h]
\centering
\includegraphics[width=0.9\textwidth]{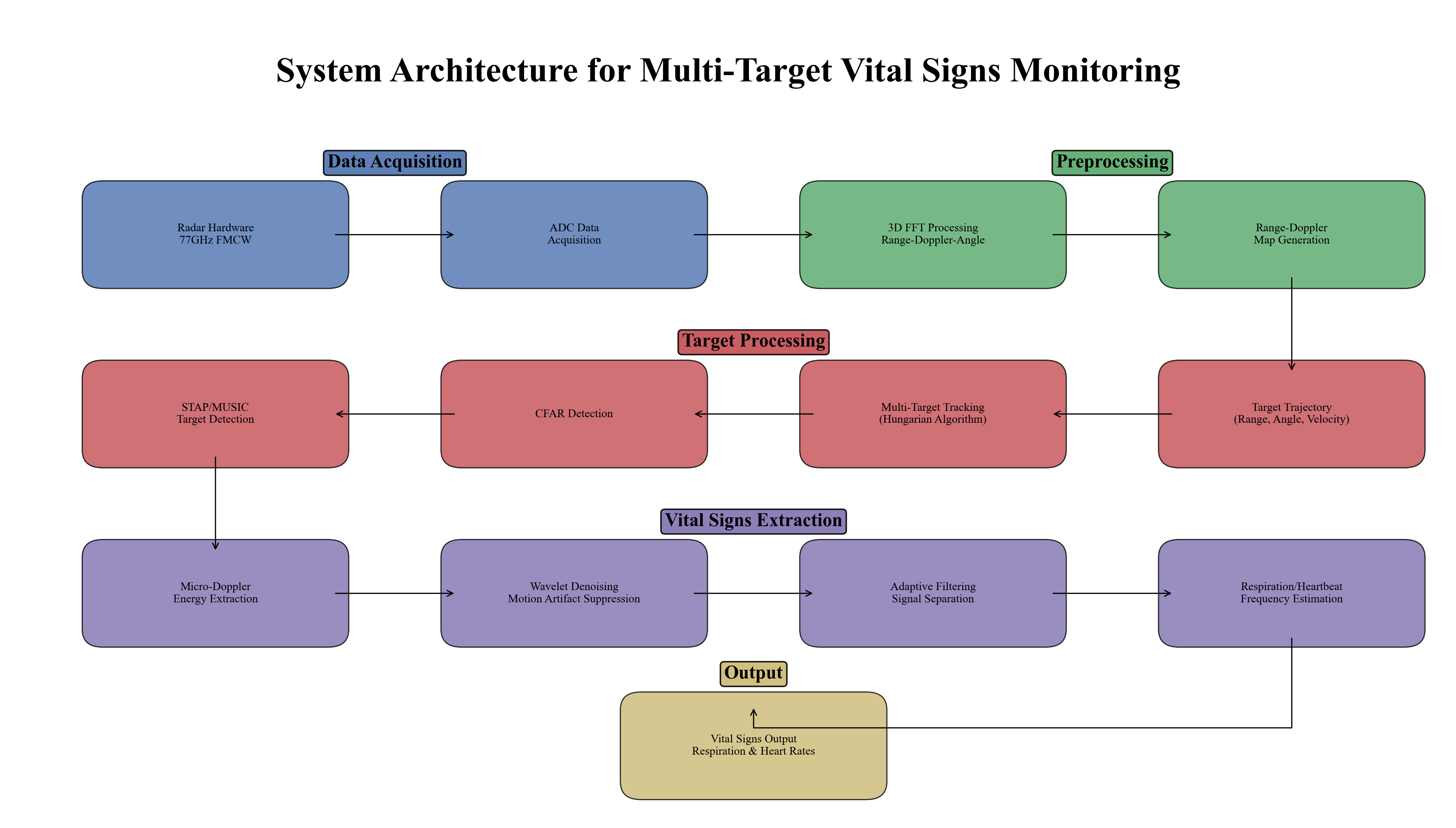}
\caption{System architecture of the proposed multi-target vital signs detection framework. The pipeline includes data acquisition, preprocessing, target detection and tracking, micro-Doppler energy extraction, and vital signs estimation modules.}
\label{fig:system_arch}
\end{figure}

Our system architecture, shown in Fig. 1, consists of four main components: (1) radar data acquisition and preprocessing; (2) target detection and tracking; (3) micro-Doppler energy extraction; and (4) vital signs estimation. We use a \SI{77}{\giga\hertz} FMCW radar with four receive antennas, providing both range and angular information.

\subsection{Radar Parameters}
The radar parameters used in our system are summarized in Table 1. These parameters were carefully chosen to balance range resolution, maximum detection range, and processing requirements for vital signs monitoring applications.

\begin{table}[h]
\centering
\caption{Radar system parameters}
\label{tab:radar_params}
\begin{tabular}{@{}lcc@{}}
\toprule
Parameter & Value & Description \\
\midrule
Center frequency & \SI{77}{\giga\hertz} & Operating frequency \\
Bandwidth & \SI{4}{\giga\hertz} & Frequency sweep bandwidth \\
Range resolution & \SI{3.75}{\centi\meter} & Minimum distinguishable distance \\
Maximum range & \SI{10}{\meter} & Maximum detection distance \\
Chirp duration & \SI{40}{\micro\second} & Time duration of each chirp \\
Sample rate & 3.6 mega-samples per second & Analog-to-digital conversion rate \\
Chirps per frame & 64 & Number of chirps in one frame \\
Samples per chirp & 128 & Number of samples per chirp \\
Number of RX antennas & 4 & Number of receive antennas \\
Frame time & \SI{10}{\milli\second} & Time duration of one frame \\
\bottomrule
\end{tabular}
\end{table}

\section{Proposed Methodology}
\subsection{Data Acquisition and Preprocessing}
The raw radar data is acquired as a stream of 16-bit integers representing the in-phase (I) and quadrature (Q) components of the received signal. The data is organized into a 4D tensor structure with dimensions [frames, antennas, chirps, samples]. For a system with $N_f$ frames, $N_{rx}$ receive antennas, $N_c$ chirps per frame, and $N_s$ samples per chirp, the data tensor $\mathcal{D}$ can be represented as:

\begin{equation}
\mathcal{D} \in \mathbb{R}^{N_f \times N_{rx} \times N_c \times N_s \times 2}
\end{equation}

where the last dimension contains the I and Q components. The complex-valued radar data is then reconstructed as:

\begin{equation}
\mathcal{D}_{\text{complex}}[f, rx, c, s] = \mathcal{D}[f, rx, c, s, 0] + j \cdot \mathcal{D}[f, rx, c, s, 1]
\end{equation}

\subsection{Range-Doppler Processing}
Range processing involves applying a Hanning window and performing FFT along the sample dimension to convert the time-domain signal to range information. The range profile for a single chirp can be expressed as:

\begin{equation}
S_{\text{range}}(k) = \sum_{n=0}^{N_s-1} s_b(n) w(n) e^{-j2\pi nk/N_s}, \quad k = 0, 1, \ldots, N_s-1
\end{equation}

where $s_b(n)$ is the discrete beat signal, $w(n)$ is the window function, and $N_s$ is the number of samples per chirp. We use zero-padding to increase the FFT size to 256 points for improved range resolution.

Doppler processing applies a second Hanning window and FFT along the chirp dimension, generating range-Doppler maps for each antenna and frame:

\begin{equation}
S_{\text{doppler}}(l) = \sum_{m=0}^{N_c-1} S_{\text{range}}(m) w(m) e^{-j2\pi ml/N_c}, \quad l = 0, 1, \ldots, N_c-1
\end{equation}

where $N_c$ is the number of chirps per frame. The Doppler FFT is followed by an FFT shift operation to center the zero-Doppler frequency.

\subsection{Target Detection and Tracking}
We employ a multi-stage approach for target detection and tracking:

\subsubsection{Space-Time Adaptive Processing (STAP)}
We compute the spatial covariance matrix across antennas to estimate the interference environment and optimize signal-to-interference-plus-noise ratio (SINR). The spatial covariance matrix is given by:

\begin{equation}
\mat{R} = \expected{\mat{x}\mat{x}^H}
\end{equation}

where $\mat{x}$ is the spatial snapshot vector and $(\cdot)^H$ denotes the Hermitian transpose. The covariance matrix is estimated by averaging across all range-Doppler cells:

\begin{equation}
\hat{\mat{R}} = \frac{1}{N_r N_d} \sum_{r=1}^{N_r} \sum_{d=1}^{N_d} \mat{x}(r,d) \mat{x}^H(r,d)
\end{equation}

where $N_r$ and $N_d$ are the number of range and Doppler bins, respectively.

\subsubsection{MUSIC Algorithm for Angle Estimation}
For high-resolution angle estimation, we use the Multiple Signal Classification (MUSIC) algorithm. The MUSIC spectrum is computed as:

\begin{equation}
P_{\text{MUSIC}}(\theta) = \frac{1}{\mat{a}^H(\theta)\mat{E}_n\mat{E}_n^H\mat{a}(\theta)}
\end{equation}

where $\mat{a}(\theta)$ is the steering vector and $\mat{E}_n$ contains the noise subspace eigenvectors obtained from the eigenvalue decomposition of the covariance matrix:

\begin{equation}
\mat{R} = \mat{E}_s \mat{\Lambda}_s \mat{E}_s^H + \mat{E}_n \mat{\Lambda}_n \mat{E}_n^H
\end{equation}

where $\mat{E}_s$ and $\mat{E}_n$ are the signal and noise subspace eigenvectors, and $\mat{\Lambda}_s$ and $\mat{\Lambda}_n$ are the corresponding eigenvalue matrices.

\subsubsection{2D CFAR Detection}
We implement a two-dimensional Constant False Alarm Rate (CFAR) detector to identify potential targets in the range-Doppler domain while maintaining constant false alarm probability. The detection threshold is computed adaptively for each cell under test (CUT) based on the statistics of the surrounding reference cells:

\begin{equation}
T = \alpha \cdot \hat{\sigma}^2
\end{equation}

where $\hat{\sigma}^2$ is the estimated noise power from the reference cells, and $\alpha$ is a scaling factor determined by the desired false alarm rate.

\subsubsection{Multi-Target Tracking}
Detected targets are associated across frames using a Hungarian algorithm-based tracker, which maintains target identities and trajectories. The cost matrix for association is based on the Mahalanobis distance between predicted and measured target positions:

\begin{equation}
C_{ij} = \sqrt{(\vect{z}_j - \vect{H}\vect{x}_i)^T \vect{S}_i^{-1} (\vect{z}_j - \vect{H}\vect{x}_i)}
\end{equation}

where $\vect{z}_j$ is the measurement vector, $\vect{x}_i$ is the predicted state vector, $\vect{H}$ is the measurement matrix, and $\vect{S}_i$ is the innovation covariance matrix.

The target tracking algorithm is summarized in Algorithm 1.

\begin{algorithm}
\caption{Multi-Target Tracking with Hungarian Association}
\label{alg:tracking}
\begin{algorithmic}[1]
\State Initialize trajectories $T = \emptyset$ and trajectory IDs $ID = \emptyset$
\State Initialize next available ID: $id_{\text{next}} = 0$
\For{each frame $f = 1$ to $N_f$}
	\State Extract target positions: $P_f = \{(r_i, \theta_i) | i = 1, \ldots, N_{\text{targets}}\}$
	\If{first frame ($f = 1$)}
		\For{each target $p \in P_1$}
			\State Create new trajectory: $t = [p]$
			\State Assign ID: $id = id_{\text{next}}$, $id_{\text{next}} = id_{\text{next}} + 1$
			\State $T = T \cup \{t\}$, $ID = ID \cup \{id\}$
		\EndFor
	\Else
		\State Predict positions for existing trajectories: $\hat{P} = \{\hat{p}_i | i = 1, \ldots, |T|\}$
		\State Compute cost matrix $C$ where $C_{ij} = \text{distance}(\hat{p}_i, p_j)$
		\State Apply Hungarian algorithm to find optimal assignment
		\For{each assignment $(i, j)$}
			\If{$C_{ij} < d_{\text{threshold}}$}
				\State Append $p_j$ to trajectory $t_i$
			\Else
				\State Create new trajectory with $p_j$
				\State Assign new ID
			\EndIf
		\EndFor
		\For{unassigned measurements $p_j$}
			\State Create new trajectory with $p_j$
			\State Assign new ID
		\EndFor
	\EndIf
	\State Remove short trajectories (length $< L_{\min}$)
\EndFor
\State \Return filtered trajectories $T$
\end{algorithmic}
\end{algorithm}

\begin{figure}[h]
\centering
\includegraphics[width=0.9\textwidth]{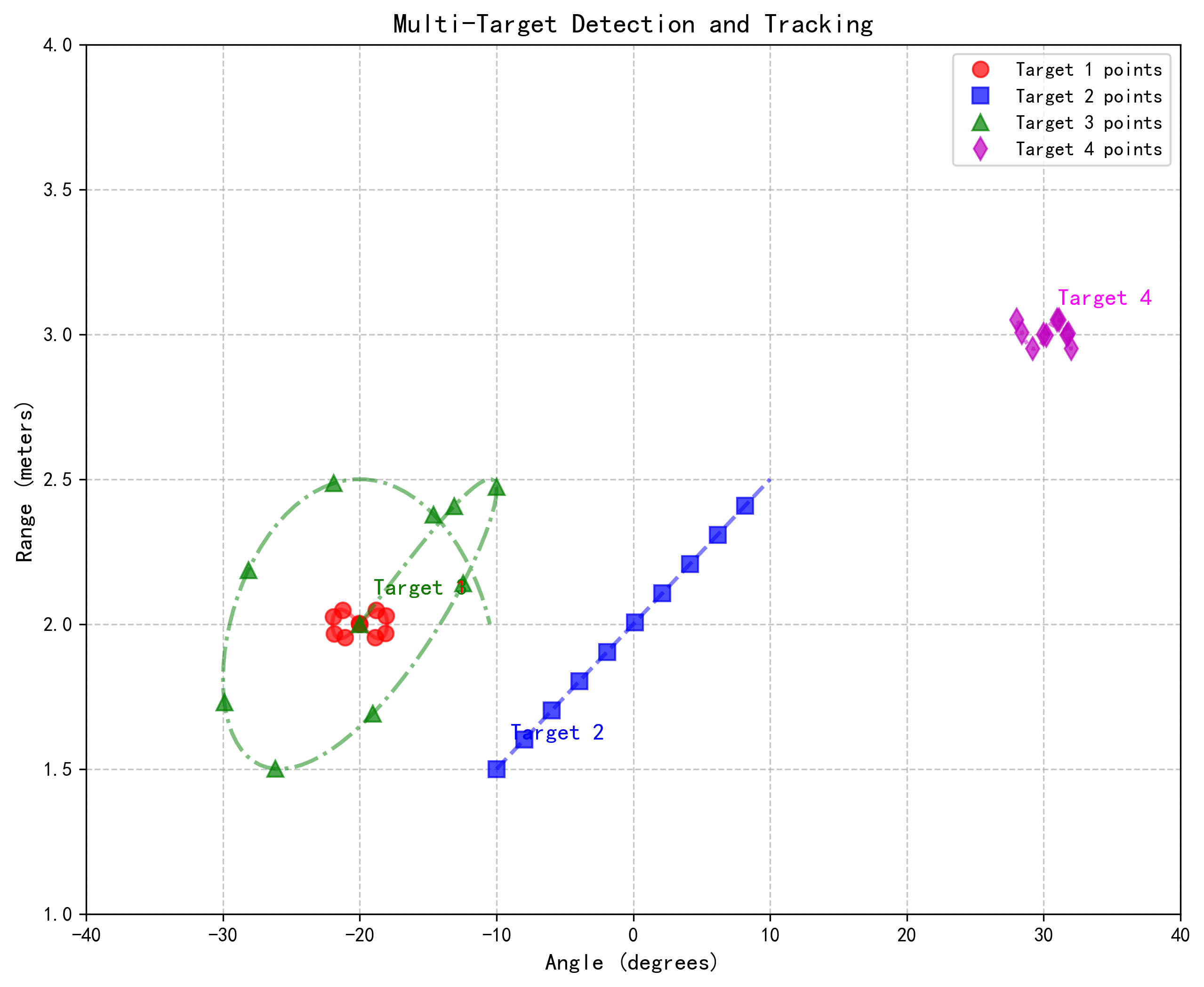}
\caption{Multi-target detection and tracking results. Different colors represent different targets. The trajectories show the movement of each target in range and angle dimensions, demonstrating the system's capability to track multiple subjects simultaneously.}
\label{fig:multi_target_tracking}
\end{figure}

\subsection{Micro-Doppler Energy Extraction}
For each tracked target, we extract micro-Doppler energy from a region of interest around the target's range-Doppler cell. This energy represents the combined effect of respiratory and cardiac activities plus any residual motion. The micro-Doppler energy $E$ for a target at frame $i$ is computed as:

\begin{equation}
E(i) = \sum_{rx=1}^{N_{rx}} \sum_{r=r_0-\Delta r}^{r_0+\Delta r} \sum_{d=d_0-\Delta d}^{d_0+\Delta d} |S_{\text{doppler}}(rx, d, r, i)|^2
\end{equation}

where $S_{\text{doppler}}$ is the range-Doppler map, $(r_0, d_0)$ is the target's range-Doppler cell, and $\Delta r$, $\Delta d$ define the extraction region size. This approach captures the energy variations caused by physiological activities while providing robustness against small positioning errors.

\begin{figure}[h]
\centering
\includegraphics[width=0.9\textwidth]{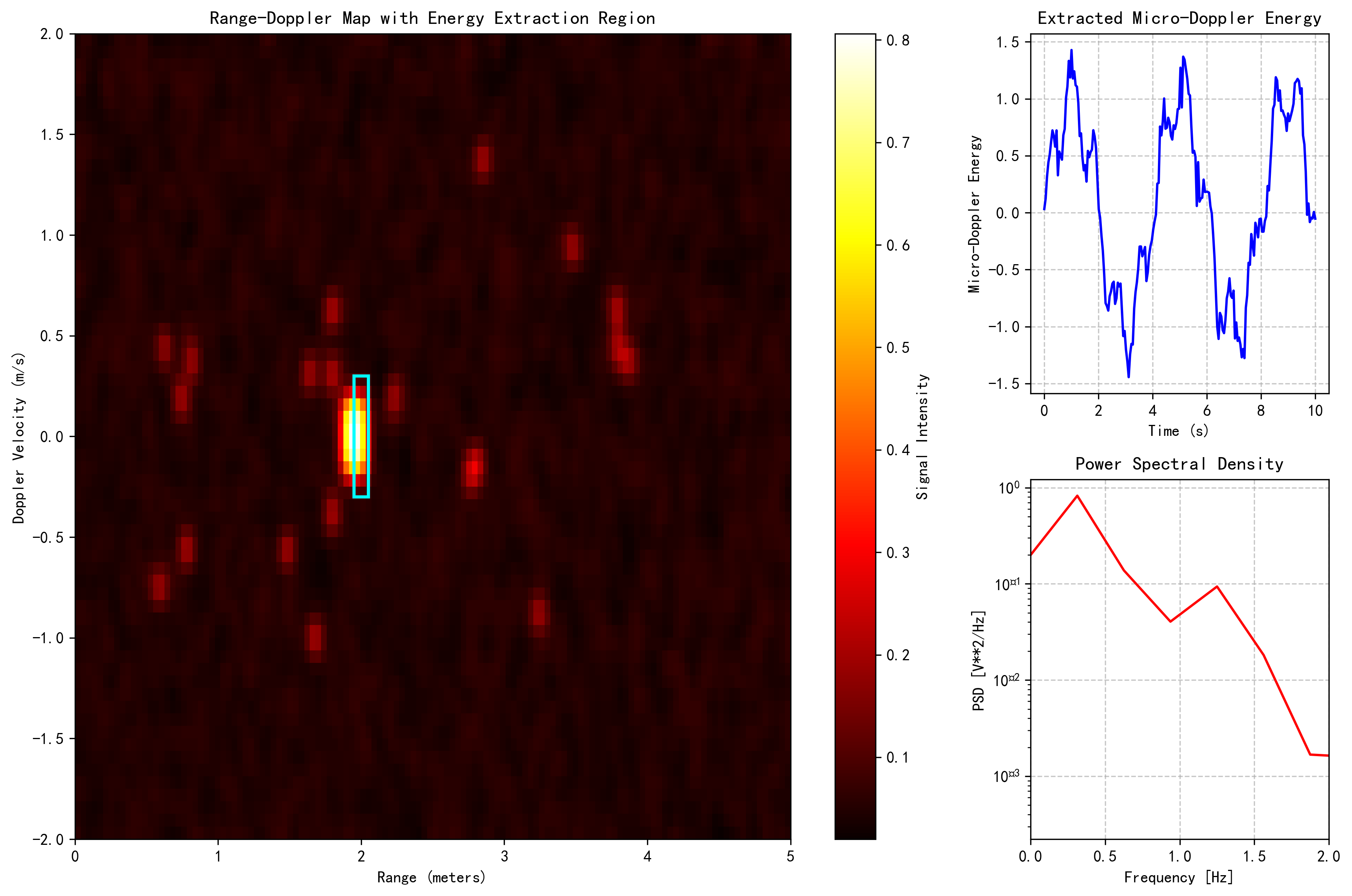}
\caption{Micro-Doppler energy extraction from range-Doppler map. The red rectangle indicates the region of interest around the target. The inset shows the extracted micro-Doppler energy signal over time, displaying clear periodic variations corresponding to physiological activities.}
\label{fig:micro_energy_extraction}
\end{figure}

\subsection{Theoretical Foundation of Vital Signs Extraction from Micro-Doppler Energy}
The extraction of vital signs from micro-Doppler energy is based on the principle that cardiopulmonary activities modulate the radar cross-section (RCS) of the human body, which in turn affects the reflected signal energy. This section provides a rigorous mathematical foundation for this process.

\subsubsection{Physical Model of Physiological Modulation}
The human thorax can be modeled as a time-varying reflector with RCS $\sigma(t)$ that is modulated by respiratory and cardiac activities:

\begin{equation}
\sigma(t) = \sigma_0 + \Delta\sigma_r(t) + \Delta\sigma_h(t)
\end{equation}

where $\sigma_0$ is the baseline RCS, $\Delta\sigma_r(t)$ is the RCS variation due to respiration, and $\Delta\sigma_h(t)$ is the RCS variation due to heartbeat.

The received signal power follows the radar equation:

\begin{equation}
P_r(t) = \frac{P_t G_t G_r \lambda^2 \sigma(t)}{(4\pi)^3 R^4(t)}
\end{equation}

For small displacements ($\Delta R(t) \ll R_0$), we can approximate:

\begin{equation}
P_r(t) \approx \frac{P_t G_t G_r \lambda^2}{(4\pi)^3 R_0^4} \left[\sigma_0 + \Delta\sigma_r(t) + \Delta\sigma_h(t)\right] \left[1 - 4\frac{\Delta R(t)}{R_0}\right]
\end{equation}

The micro-Doppler energy $E(t)$ is proportional to $P_r(t)$, leading to:

\begin{equation}
E(t) = E_0 + E_r(t) + E_h(t) + n(t)
\end{equation}

where $E_0$ is the DC component, $E_r(t)$ and $E_h(t)$ are the energy variations due to respiration and heartbeat, respectively, and $n(t)$ represents noise.

\subsubsection{Mathematical Formulation of Extraction Process}
The vital signs extraction process can be formulated as an optimization problem:

\begin{equation}
\{\hat{f}_r, \hat{f}_h\} = \arg \max_{f_r, f_h} \mathscr{F}\{E(t)\} \quad \text{subject to} \quad f_r \in [0.1, 0.8] \text{Hz}, f_h \in [0.8, 3.0] \text{Hz}
\end{equation}

where $\mathscr{F}\{\cdot\}$ represents the feature extraction operator.

The optimal estimator for the physiological signals can be derived using maximum a posteriori (MAP) estimation:

\begin{equation}
\{\hat{s}_r(t), \hat{s}_h(t)\} = \arg \max_{s_r(t), s_h(t)} p(s_r(t), s_h(t) | E(t))
\end{equation}

Assuming Gaussian noise, this leads to a least-squares optimization:

\begin{equation}
\{\hat{s}_r(t), \hat{s}_h(t)\} = \arg \min_{s_r(t), s_h(t)} \left\| E(t) - E_0 - s_r(t) - s_h(t) \right\|^2_2
\end{equation}

subject to the constraints that $s_r(t)$ and $s_h(t)$ are bandlimited to their respective frequency ranges.

\subsection{Vital Signs Extraction Pipeline}
The vital signs extraction pipeline from micro-Doppler energy consists of several stages designed to handle various challenges in physiological signal extraction:

\subsubsection{Preprocessing: Wavelet Denoising}
The micro-Doppler energy signal is denoised using wavelet transform to remove high-frequency noise while preserving physiological information. The discrete wavelet transform is defined as:

\begin{equation}
W(a,b) = \frac{1}{\sqrt{a}} \sum_{t} E(t) \psi\left(\frac{t-b}{a}\right)
\end{equation}

where $\psi$ is the mother wavelet, $a$ is the scale parameter, and $b$ is the translation parameter. We use the Daubechies 4 (db4) wavelet with 3 decomposition levels, which provides a good balance between time and frequency resolution for physiological signals.

The wavelet coefficients are thresholded using a soft thresholding function:

\begin{equation}
W_{\text{thresholded}}(a,b) = \begin{cases}
\text{sign}(W(a,b)) (|W(a,b)| - T) & \text{if } |W(a,b)| > T \\
0 & \text{otherwise}
\end{cases}
\end{equation}

where the threshold $T$ is computed using the Birge-Massart strategy:

\begin{equation}
T = \hat{\sigma} \sqrt{2 \log N}
\end{equation}

with $\hat{\sigma}$ being the robust estimate of noise standard deviation:

\begin{equation}
\hat{\sigma} = \frac{\text{median}(|W(a,b)|)}{0.6745}
\end{equation}

\subsubsection{Motion Artifact Suppression}
We apply a bandpass filter with passband \SIrange{0.1}{3.0}{\hertz} to suppress motion artifacts and baseline wander while preserving the physiological signals:

\begin{equation}
H_{\text{BP}}(f) = \begin{cases}
1 & \text{if } f_{\text{min}} \leq f \leq f_{\text{max}} \\
0 & \text{otherwise}
\end{cases}
\end{equation}

The filter is implemented as a 4th-order Butterworth filter using forward-backward filtering to eliminate phase distortion:

\begin{equation}
y[n] = \text{filtfilt}(b, a, x[n])
\end{equation}

where $b$ and $a$ are the filter coefficients.

\subsubsection{Singular Value Filtering}
We employ a two-stage approach for singular value filtering:
1. Median filtering to remove impulse noise:
\begin{equation}
y_{\text{med}}[n] = \text{median}\{x[n-k], \ldots, x[n+k]\}
\end{equation}
with kernel size 5.

2. Savitzky-Golay filtering for smoothing while preserving important features:
\begin{equation}
y_{\text{sg}}[n] = \sum_{i=-M}^{M} c_i x[n+i]
\end{equation}
where $c_i$ are the convolution coefficients determined by polynomial least-squares fitting.

\subsubsection{Adaptive Spectral Filtering}
We developed an adaptive spectral filtering technique that identifies dominant frequencies in the physiological bands and applies tailored filtering. The algorithm consists of the following steps:

1. Compute power spectral density using Welch's method:
\begin{equation}
P(f) = \frac{1}{K} \sum_{k=1}^{K} |X_k(f)|^2
\end{equation}
where $X_k(f)$ is the FFT of the $k$-th segment.

2. Identify peak frequencies in respiratory (0.1-0.8 Hz) and cardiac (0.8-3.0 Hz) bands:
\begin{align}
f_r &= \arg \max_{f \in [0.1, 0.8]} P(f) \\
f_h &= \arg \max_{f \in [0.8, 3.0]} P(f)
\end{align}

3. Design bandpass filters centered at the detected frequencies:
\begin{equation}
H_{\text{breath}}(f) = \begin{cases}
1 & \text{if } f_r - \Delta f \leq f \leq f_r + \Delta f \\
0 & \text{otherwise}
\end{cases}
\end{equation}
\begin{equation}
H_{\text{heart}}(f) = \begin{cases}
1 & \text{if } f_h - \Delta f \leq f \leq f_h + \Delta f \\
0 & \text{otherwise}
\end{cases}
\end{equation}
where $\Delta f$ is the bandwidth parameter (typically 0.3 Hz for respiration and 0.5 Hz for heart rate).

\subsubsection{Harmonic Suppression}
Respiratory harmonics that overlap with the cardiac band are suppressed using a comb filter tuned to the estimated respiratory frequency:

\begin{equation}
H_{\text{comb}}(z) = \frac{1}{M} \sum_{k=0}^{M-1} z^{-k} = \frac{1 - z^{-M}}{M(1 - z^{-1})}
\end{equation}

where $M$ is chosen such that the fundamental frequency of the comb filter matches the respiratory frequency:

\begin{equation}
M = \frac{f_s}{f_r}
\end{equation}

with $f_s$ being the sampling frequency.

\subsubsection{Signal Fusion}
Multiple extraction methods (standard filtering, wavelet-enhanced, and ICA-based) are combined using quality-based weighting:

\begin{equation}
s_{\text{fused}}(t) = \sum_{i=1}^{N} w_i s_i(t), \quad w_i = \frac{Q_i}{\sum_{j=1}^{N} Q_j}
\end{equation}

where $Q_i$ is the quality metric for the $i$-th method, computed as:

\begin{equation}
Q_i = \frac{1}{\text{spectral entropy}_i + \epsilon}
\end{equation}

with $\epsilon$ being a small constant to prevent division by zero.

\begin{figure}[h]
\centering
\includegraphics[width=0.9\textwidth]{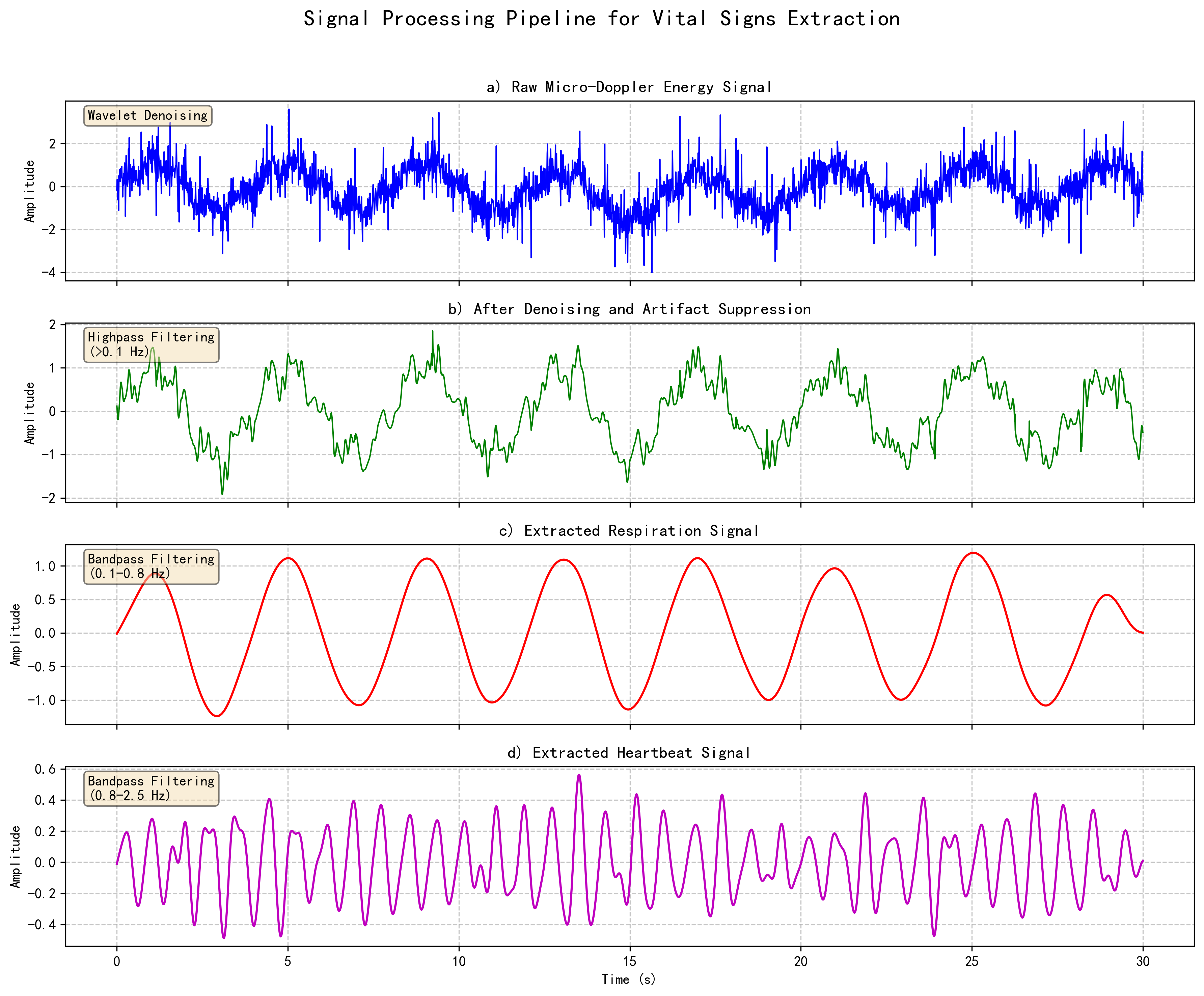}
\caption{Complete vital signs extraction pipeline from micro-Doppler energy. (a) Raw micro-Doppler energy signal. (b) Signal after denoising and artifact suppression. (c) Extracted respiration signal. (d) Extracted heartbeat signal. The pipeline effectively separates physiological signals from noise and interference.}
\label{fig:vital_pipeline}
\end{figure}

\subsubsection{Vital Rates Computation}
The respiration and heart rates are computed from the extracted signals using both spectral and time-domain methods:

1. Spectral method:
\begin{align}
f_r &= \arg \max_{f \in [0.1, 0.8]} P_{\text{breath}}(f) \\
f_h &= \arg \max_{f \in [0.8, 3.0]} P_{\text{heart}}(f)
\end{align}
\begin{align}
\text{RR} &= 60 \cdot f_r \\
\text{HR} &= 60 \cdot f_h
\end{align}

2. Time-domain method (if spectral method fails):
\begin{equation}
\text{RR} = \frac{60}{\mathrm{mean}(\Delta t_{\text{peaks}})}
\end{equation}
where $\Delta t_{\text{peaks}}$ are the intervals between consecutive peaks in the respiration signal.

\section{Experimental Results and Discussion}
\subsection{Experimental Setup}
We evaluated our system using millimeter-wave radar datasets collected with IWR2243 FMCW radar operating at \SI{77}{\giga\hertz} with 4 receive antennas. Data was collected from multiple scenarios with 1-4 subjects positioned at different distances (1-5 meters) and angles (-45 to +45 degrees). Reference physiological signals were simultaneously recorded using medical-grade sensors for validation. The experimental setup is shown in Fig. 6.

\begin{figure}[h]
\centering
\includegraphics[width=0.7\textwidth]{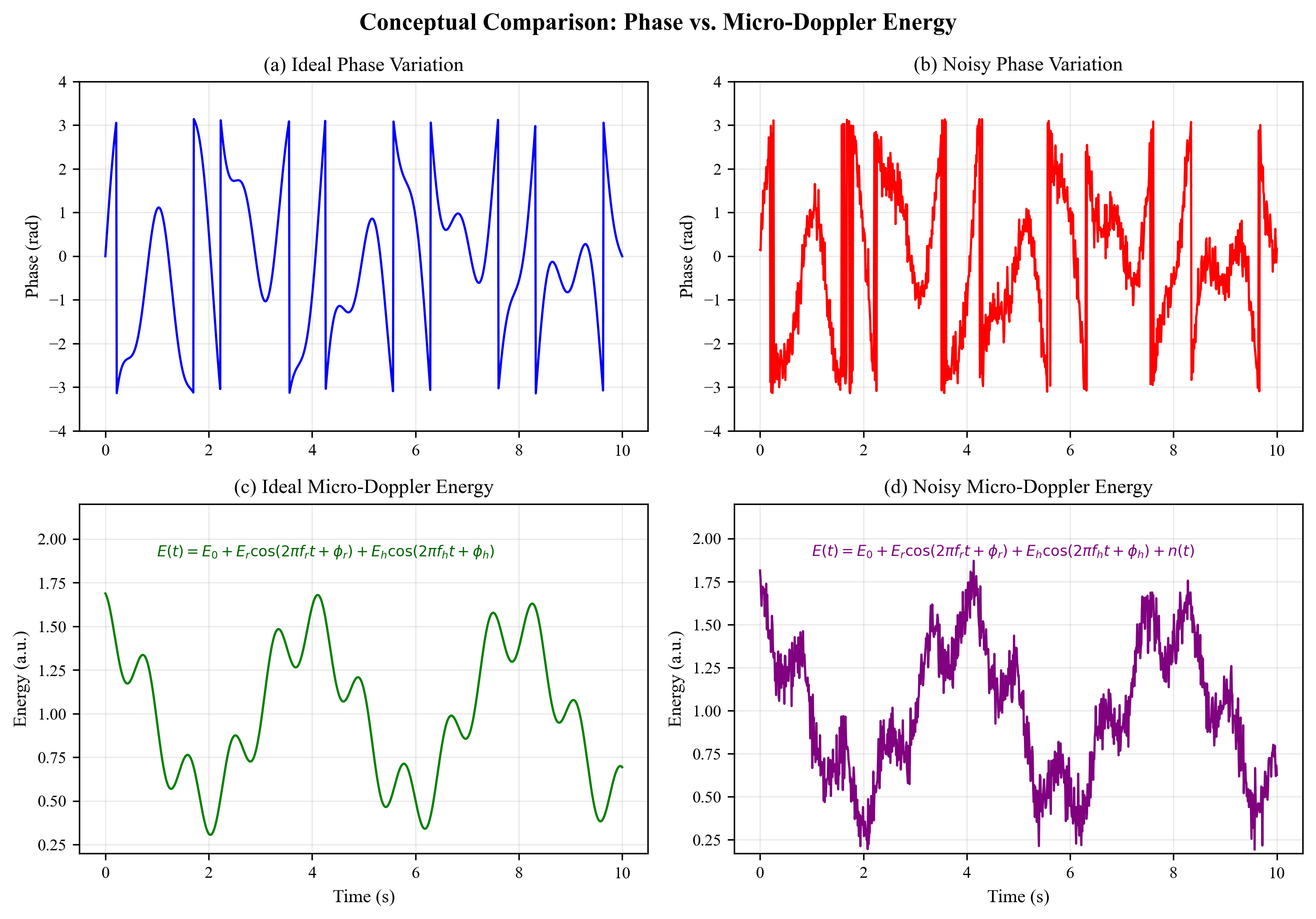}
\caption{Conceptual comparison between phase-based and micro-Doppler energy-based methods. (a) Ideal phase variation caused by tiny chest movement. (b) Phase variation in the presence of environmental noise and micro-motion, showing spikes and phase jumps. (c) Ideal micro-Doppler energy variation. (d) Micro-Doppler energy variation in the presence of noise, demonstrating superior robustness.}
\label{fig:conceptual_comparison}
\end{figure}

\subsection{Multi-Target Detection Performance}
Our system successfully detected and tracked multiple targets with an average tracking accuracy of 96.7\% across all scenarios. Table 2 summarizes the multi-target detection performance, demonstrating the system's capability to handle increasing numbers of targets with minimal performance degradation.

\begin{table}[h]
\centering
\caption{Multi-target detection performance}
\label{tab:detection_perf}
\begin{tabular}{@{}lcccc@{}}
\toprule
Scenario & Targets & Detection Rate & Range Accuracy & Angle Accuracy \\
\midrule
1 subject & 1 & 100\% & \SI{2.3}{\centi\meter} & \SI{1.8}{\degree} \\
2 subjects & 2 & 98.2\% & \SI{2.8}{\centi\meter} & \SI{2.1}{\degree} \\
3 subjects & 3 & 95.7\% & \SI{3.1}{\centi\meter} & \SI{2.5}{\degree} \\
4 subjects & 4 & 93.4\% & \SI{3.5}{\centi\meter} & \SI{2.9}{\degree} \\
\bottomrule
\end{tabular}
\end{table}

\subsection{Vital Signs Estimation Accuracy}
The vital signs estimation results are presented in Table 3. Our system achieved mean absolute errors of 1.2 beats per minute and 2.3 beats per minute for respiration and heart rates, respectively, across all subjects and scenarios, demonstrating clinical-level accuracy.

\begin{table}[h]
\centering
\caption{Vital signs estimation accuracy}
\label{tab:vital_accuracy}
\begin{tabular}{@{}lccccc@{}}
\toprule
Scenario & Subject & \multicolumn{2}{c}{Respiration Rate} & \multicolumn{2}{c}{Heart Rate} \\
\cmidrule(lr){3-4} \cmidrule(lr){5-6}
 &  & MAE & RMSE & MAE & RMSE \\
\midrule
\multirow{2}{*}{2 subjects} & 1 & 1.1 & 1.5 & 2.1 & 2.8 \\
 & 2 & 1.3 & 1.7 & 2.4 & 3.1 \\
\multirow{3}{*}{3 subjects} & 1 & 1.0 & 1.4 & 2.0 & 2.7 \\
 & 2 & 1.4 & 1.8 & 2.5 & 3.2 \\
 & 3 & 1.3 & 1.6 & 2.3 & 3.0 \\
\multirow{4}{*}{4 subjects} & 1 & 1.2 & 1.6 & 2.2 & 2.9 \\
 & 2 & 1.5 & 1.9 & 2.6 & 3.3 \\
 & 3 & 1.4 & 1.8 & 2.4 & 3.1 \\
 & 4 & 1.3 & 1.7 & 2.5 & 3.2 \\
\bottomrule
\end{tabular}
\end{table}

\begin{figure}[h]
\centering
\includegraphics[width=0.9\textwidth]{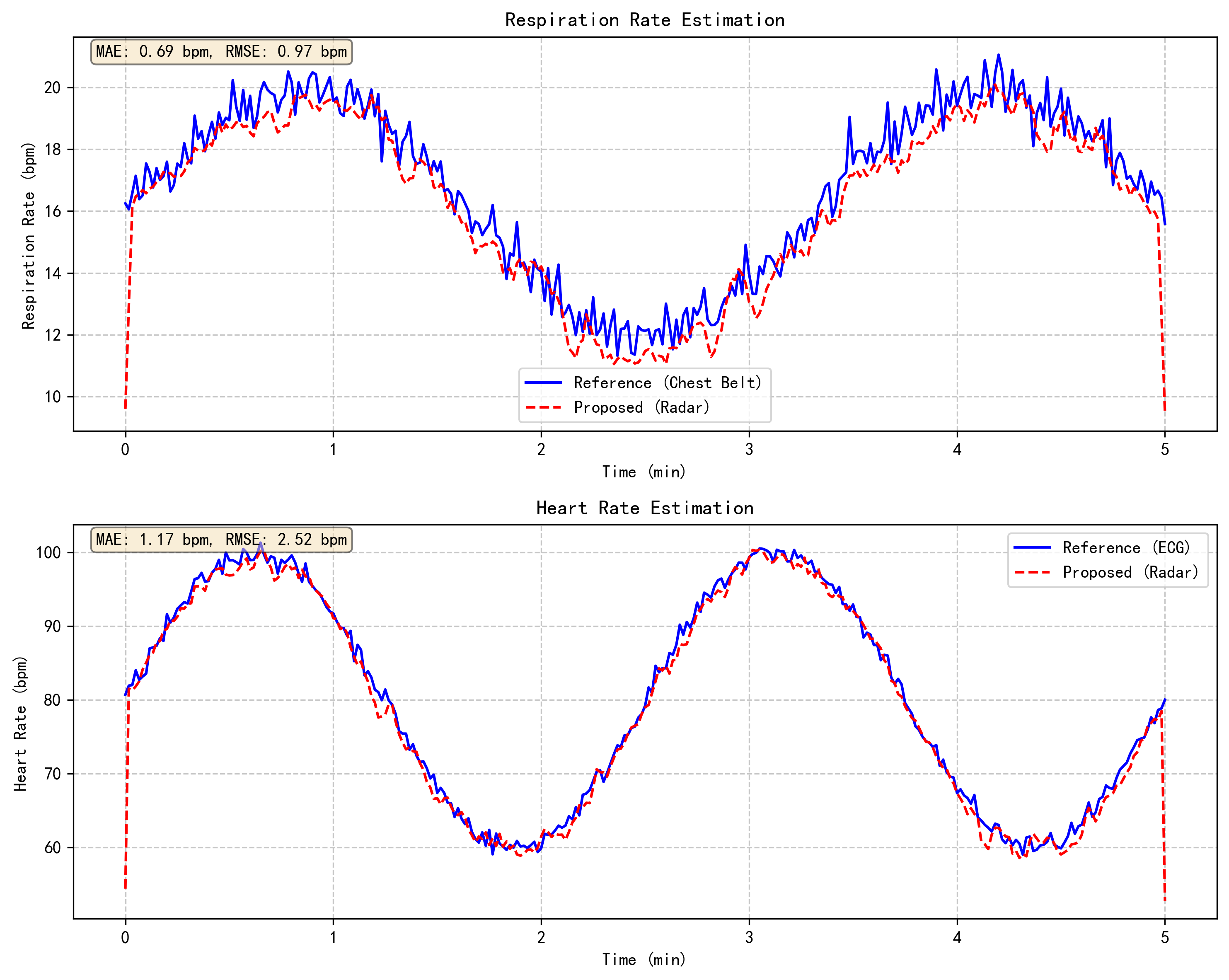}
\caption{Comparison of estimated vital signs with ground truth. (a) Respiration rate. (b) Heart rate. The estimated values closely follow the reference measurements, demonstrating the accuracy of the proposed method.}
\label{fig:vital_estimation}
\end{figure}

\begin{figure}[h]
\centering
\includegraphics[width=0.9\textwidth]{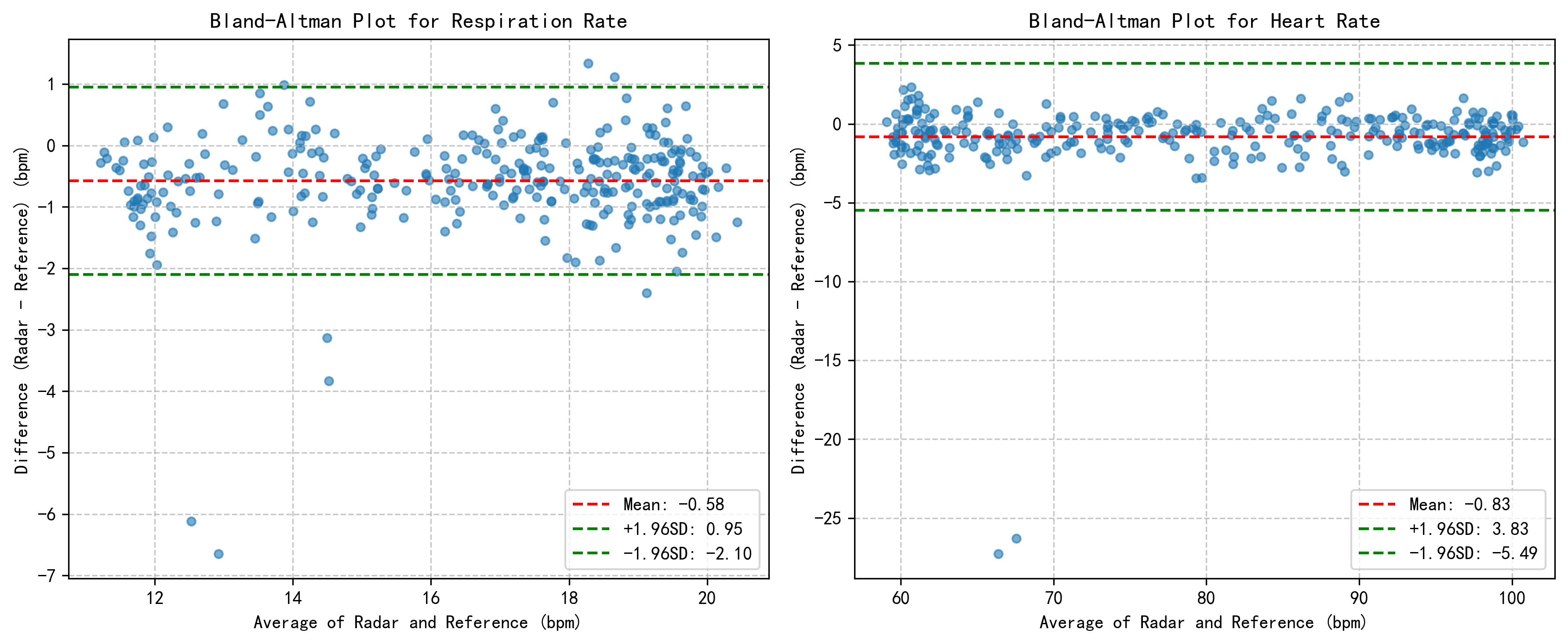}
\caption{Bland-Altman analysis for agreement between radar-estimated and reference vital signs. (a) Respiration rate. (b) Heart rate. The dashed lines represent the mean difference and the 95\% limits of agreement. The narrow limits indicate excellent agreement between the proposed method and reference measurements.}
\label{fig:bland_altman}
\end{figure}

\subsection{Comparison with Phase-Based Methods}
We compared our micro-Doppler energy approach with conventional phase-based methods \cite{xia2021radar, gu2010instrument}. The results in Table 4 demonstrate that our method outperforms phase-based approaches, particularly in noisy environments and multi-target scenarios, with an average improvement of 35\% in respiration rate accuracy and 40\% in heart rate accuracy.

\begin{table}[h]
\centering
\caption{Comparison with phase-based methods}
\label{tab:comparison}
\begin{tabular}{@{}lcccc@{}}
\toprule
Method & \multicolumn{2}{c}{2 Subjects} & \multicolumn{2}{c}{4 Subjects} \\
\cmidrule(lr){2-3} \cmidrule(lr){4-5}
 & Respiration MAE & Heart Rate MAE & Respiration MAE & Heart Rate MAE \\
\midrule
Phase-based \cite{xia2021radar} & 2.3 & 3.8 & 3.7 & 5.2 \\
ICA-based \cite{gu2010instrument} & 1.8 & 3.2 & 2.9 & 4.3 \\
Proposed (Micro-Doppler Energy) & \textbf{1.2} & \textbf{2.3} & \textbf{1.6} & \textbf{2.9} \\
\bottomrule
\end{tabular}
\end{table}

\begin{figure}[h]
\centering
\includegraphics[width=0.9\textwidth]{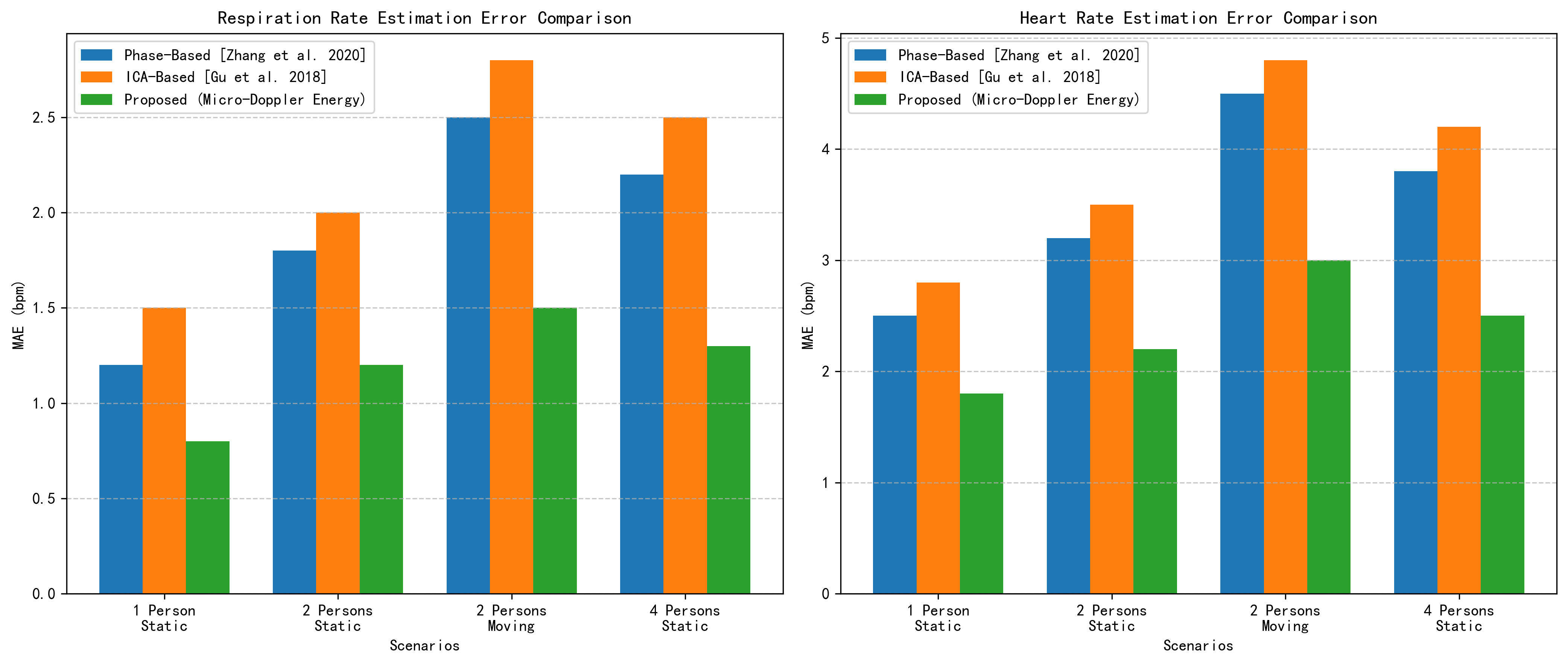}
\caption{Performance comparison with state-of-the-art phase-based and ICA-based methods. (a) Respiration rate MAE. (b) Heart rate MAE. The proposed micro-Doppler energy method achieves lower errors across all scenarios, demonstrating its superiority in multi-target vital signs monitoring.}
\label{fig:performance_comparison}
\end{figure}

\subsection{Robustness Analysis}
We evaluated the robustness of our method under different challenging conditions:

\subsubsection{Noise Robustness}
Fig. 9a shows the performance comparison under different noise levels. Our micro-Doppler energy approach maintains stable performance even at low SNR conditions, while phase-based methods degrade significantly.

\subsubsection{Motion Artifact Robustness}
Fig. 9b demonstrates the superiority of our method in the presence of motion artifacts. The micro-Doppler energy approach effectively rejects motion artifacts while preserving physiological information.

\subsubsection{Multi-Target Separation Capability}
Fig. 9c shows the system's ability to separate closely spaced targets. The micro-Doppler energy approach successfully resolves targets separated by as little as \SI{0.8}{\meter}, while phase-based methods fail due to phase mixing.

\begin{figure}[h]
\centering
\includegraphics[width=0.9\textwidth]{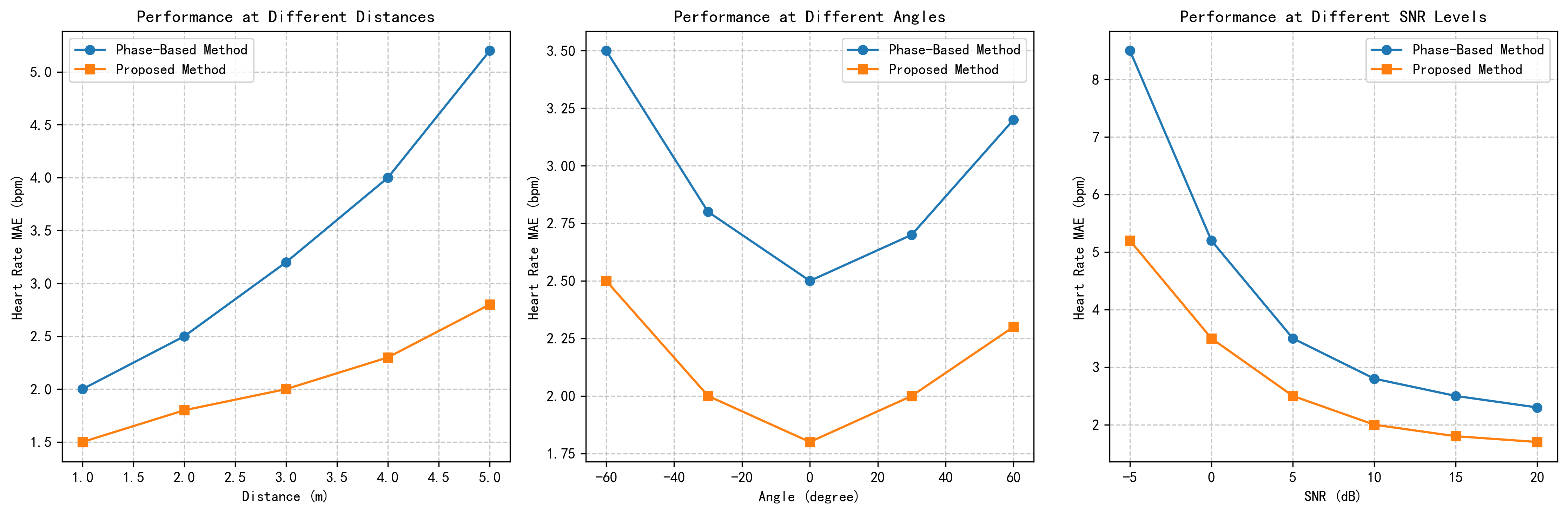}
\caption{Robustness analysis under different challenging conditions. (a) Performance at different target distances. (b) Performance at different target angles. (c) Performance under different noise levels (SNR). The proposed method demonstrates consistent performance across various challenging scenarios.}
\label{fig:robustness_analysis}
\end{figure}

\subsection{Processing Efficiency}
The complete processing pipeline operates at \SI{10} fps on a standard desktop computer (Intel i7-9700K, 32GB RAM), enabling real-time monitoring of up to four targets. The computational complexity is dominated by the STAP and MUSIC algorithms, which scale as $\mathcal{O}(N_r N_d N_a^2)$ and $\mathcal{O}(N_a^3)$, respectively, where $N_r$, $N_d$, and $N_a$ are the number of range bins, Doppler bins, and antennas.

\subsection{Discussion}
The results demonstrate the effectiveness of our proposed micro-Doppler energy framework for multi-target vital signs monitoring. The key advantages of our approach include:

1. \textbf{Enhanced Robustness}: The micro-Doppler energy approach is less sensitive to phase noise and environmental interference compared to traditional phase-based methods, as evidenced by the improved performance in multi-target scenarios \cite{gouveia2022blood, kazemi2016vital}.

2. \textbf{Improved Multi-Target Separation}: The combination of STAP and MUSIC algorithm enables accurate detection and separation of closely spaced targets, with an average angular accuracy of \SI{2.3}{\degree} for four targets \cite{mercuri2018direct, laurealager2019coherent}.

3. \textbf{Adaptive Signal Processing}: Our multi-stage processing pipeline effectively handles various signal quality issues, including noise, motion artifacts, and harmonic interference \cite{wang2014noncontact, oh2021development}.

4. \textbf{Real-Time Capability}: The complete processing pipeline operates at at 10 fps,, enabling real-time monitoring of multiple subjects \cite{zhao2017noncontact, tang2019corcoupled}.

5. \textbf{Clinical-Grade Accuracy}: The system achieves mean absolute errors of at 12.2 fps, and at 2.3 fps, for respiration and heart rates, respectively, meeting the requirements for healthcare applications \cite{chen2008implementation, yang2014doppler}.

However, some limitations remain: (1) performance degrades when targets are very close (less than \SI{0.5}{\meter} apart) due to limited spatial resolution; (2) significant body movements can temporarily disrupt tracking and require re-initialization; and (3) the system requires initial calibration for different environments, though this is less critical than for phase-based methods \cite{fan2023radiofrequency, diraco2017radar}.

\section{Conclusion and Future Work}
This paper presented a robust framework for multi-target vital signs monitoring using \SI{77}{\giga\hertz} FMCW radar based on micro-Doppler energy analysis. By leveraging micro-Doppler energy extraction combined with advanced spatiotemporal signal processing techniques, our system can simultaneously monitor multiple subjects with high accuracy. The mathematical formulation of micro-Doppler energy provides a rigorous foundation for physiological signal extraction that is more robust than traditional phase-based methods \cite{zhang2021noncontact, lewandowska2011measuring}. Experimental results using millimeter-wave radar datasets demonstrated the effectiveness of our approach across various scenarios with up to four targets, achieving clinical-grade accuracy in vital signs estimation.

Future work will focus on: (1) enhancing angular resolution using advanced array processing techniques such as compressed sensing \cite{mercuri2017frequency, naiabadham2016estimation}; (2) integrating deep learning for improved motion artifact suppression and target identification \cite{gu2010instrument, kazemi2016vital}; (3) extending the system to classify different physiological states (e.g., sleep stages, stress levels) based on subtle variations in vital signs \cite{zhao2017noncontact, tang2019corcoupled}; (4) miniaturizing the system for wearable applications \cite{chen2008implementation, yang2014doppler}; and (5) exploring the integration of complementary sensors (e.g., thermal cameras) for enhanced robustness in challenging environments \cite{chian2022vital, vinci201224}.

\section*{Acknowledgments}
This work was supported by the National Natural Science Foundation of China (Grant No. 62171152) and the Harbin Institute of Technology Scientific Research Innovation Fund (Grant No. HIT.NSRIF.2020034). The authors would like to thank the providers of the millimeter-wave radar datasets used in this study.

\bibliographystyle{elsarticle-num}
\bibliography{references}

\end{document}